# Assessment of cortical reorganization and preserved function in phantom limb pain: a methodological perspective

**Jamila Andoh[1], Christopher Milde[1,2], Martin Diers[1,3], Robin Bekrater-Bodmann[1], Jörg Trojan[1,2], Xaver Fuchs[1,4], Susanne Becker[1], Simon Desch[1], Herta Flor[1]**

[1] Department of Cognitive and Clinical Neuroscience, Central Institute of Mental Health, Medical Faculty Mannheim, Heidelberg University, Mannheim, Germany.

[2] Department of Psychology, University of Koblenz-Landau, Landau, Germany.

[3] Department of Psychosomatic Medicine and Psychotherapy, LWL University Hospital, Ruhr-University Bochum, Bochum, Germany.

[4] Biopsychology & Cognitive Neuroscience, Faculty of Psychology & Sports Science, Bielefeld University, Bielefeld, Germany.

Correspondence should be addressed to

Jamila Andoh, PhD or Herta Flor, PhD

Department of Cognitive and Clinical Neuroscience,

Central Institute of Mental Health,

J5,

D-68159 Mannheim, Germany,

Phone: +49 621 1703 6302

Fax: +49 621 1703 6305

jamila.andoh@zi-mannheim.de, herta.flor@zi-mannheim.de

**Abstract**

Phantom limb pain (PLP) has been associated with both the reorganization of the somatotopic map in primary somatosensory cortex (S1) and preserved S1 function. Here we assessed the nature of the information (sensory, motor) that reaches S1 and methodological differences in the assessment of cortical representations that might explain these findings.

We used functional magnetic resonance imaging during a virtual hand motor task, analogous to the classical mirror box task, in amputees with and without PLP and in matched controls. We assessed the relationship between task-related activation maxima and PLP intensity in S1 versus motor (M1) maps. We also measured cortical distances between the location of activation maxima using region of interest (ROI), defined from individual or group analyses.

Amputees showed significantly increased activity in M1 and S1 compared to controls, which was not significantly related to PLP intensity. The location of activation maxima differed between groups depending on M1 and S1 maps. Neural activity was positively related to PLP only in amputees with pain and only if a group-defined ROI was chosen. These findings show that the analysis of changes in S1 versus M1 yields different results, possibly indicating different pain-related processes.

**Introduction**

Extensive research has shown that increased use or sensory stimulation of limbs results in an enlarged cortical representation of this limb in contralateral primary somatosensory (S1) and motor (M1) cortices. In contrast, reduced use of a limb results in a decrease of its S1 representation, and often in an expansion of the representation area of adjacent body parts [1-3]. In addition, injury, such as the amputation of a limb, has been found to lead to changes in the configuration of the cortical map with the amount of input from adjacent areas being related to phantom limb pain [4,5]. Using an active motor task involving phantom movements or motor imagery for amputees who could not move the phantom, a positive relationship between the magnitude of PLP and the intensity (peak level) of brain activation in the S1 region representing the phantom hand was reported, suggesting preserved function within S1 [6].

Maladaptive plasticity was first shown by examining neural activity related to passive sensory stimulation applied over the mouth and the intact hand using magnetoencephalography [4,5]. Flor et al. (1995) applied tactile stimuli over the lips and the fingers and showed a displacement of the lip towards the hand somatosensory area, which was related to PLP, such that the closer the lip was to the hand somatosensory area, the more intense the PLP. Such findings were also replicated with tasks in the motor domain. Using lip pursing and imagined movements of the phantom hand, an expansion of neural activity from the lip into the hand areas of both S1 and M1 was found in amputees with PLP and this shift was positively correlated with PLP[7]. Diers et al. showed that mirrored movement of the intact hand failed to activate the phantom cortex in S1 and M1 in amputees with phantom pain, whereas normal activation was found in amputees without PLP [8,9]. The more activation in the phantom cortex, the less phantom limb pain was observed. These findings seem in opposition from those of Makin et al., (2013) who found a positive association between phantom limb pain and activation in the S1M1 phantom cortex suggesting that the hand motor representation remains active after amputation. In Makin et al., a cortical representation of the phantom hand was defined using a conjunction analysis to reveal shared neural activity between execution or imagery of phantom movements in 18 amputees (16 with, two without PLP and 11 persons with a congenital upper limb deficiency) and intact hand movements in 22 two-handers in brain areas encompassing central and postcentral sulci (i.e., S1 and M1). The neural activity the authors correlated with PLP was thus based on joint activation maxima of all groups irrespective of whether there had been a change in the hand representations in M1 and S1 in the PLP group [1,7,10]. In this case, the use of a conjunction analysis might not have captured the current (post amputation) hand representation but the original preamputation location.

An additional important factor relates to the task being used. For instance, Makin et al., (2013) combined neural activity resulting from execution and imagery of movements of the phantom hand, whereas Lotze et al. (2001) and MacIver et al. (2008) used imagery of phantom movements only and Diers et al. (2010) used mirrored movements. Both processes, motor imagery and motor execution, have been shown to have a different neural representations [11] and might therefore show a different relationship with PLP.

Such differences are accentuated by the fact that execution of phantom movements is slower compared with imagery of phantom movements, possibly due to a lack of expected sensory feedback, while there is no feedback expectation during imagination [12]. In addition, presence of phantom pain occurs more often during motor execution than during motor imagery [12], which might lead to different patterns of neural activity between amputees with and without PLP related to the neural activation of the phantom percept.

Moreover, execution of phantom movements is often accompanied by activity in the muscles of the residual limb [13], therefore activity in the motor cortex might result from associated muscle movements from the residual limb, especially if the participants are trained to use these muscles during phantom movement as reported in Makin et al. (2013). The role of the residual limb during the execution of phantom hand movements has also been shown using transcranial magnetic stimulation applied over the representation of the amputated hand [14]. This evoked not only phantom movements but also contractions in the muscles of the residual limb, raising the question of whether the phantom movements stimulated the proximal arm area of the sensorimotor cortex [14].

Such "confound" of neural activity in the sensorimotor cortex during phantom movements might not only arise from the residual muscles but also from other body parts such as the lips, elbow or feet [15], possibly resulting from compensatory behaviors [16]. For instance, movements of the lips or the feet increased neural activity in the missing hand representation in congenital one-handers [16]. Similarly, movements of the existing upper arms in an individual with congenitally absent limbs [17] and movement of the lips and elbow in amputees induced activity in the deprived hand area [15]. Therefore, factors such as residual limb activity and compensatory use might induce sensorimotor plasticity, affecting the assessment of the neural representation of the missing hand. In the presence of phantom pain, these factors might even play a bigger role.

Since a number of methodological factors influence the sensory and motor representations observed in amputees and their relation to phantom limb pain, we used fMRI combined with a virtual mirror task, which was performed identically by all amputees and matched two-handed controls. In this task, all participants were moving their intact hand while seeing synchronous movements of the phantom hand (or mirrored hand in controls) in a virtual environment. We also trained the amputees not to move the residual limb, which was verified by measuring motor activity in the residual limb using electromyography in a subsample of the amputees.

We examined task-related neural activity separately in the primary somatosensory (S1), motor (M1) and sensorimotor cortices (S1M1, analogous to Makin et al., 2013) in the deafferented/matched hemisphere. We also assessed differences in neural activity between amputees with and without phantom pain and healthy controls, and their relationship to phantom pain using cortical distance measures and we employed both averaged and group-specific regions of interest. Our hypotheses stated that the neural activity resulting from the virtual mirror task might differentially involve S1 and M1 in amputees with and without PLP, and controls. In addition, we expected to find different relationships with PLP depending on which aproach is being used, i.e. using activation maxima (either individually defined or based on group activation), or if one examines motor or somatosensory cortices, or if ones calculates distance measures.

## Methods

### *Participants*

Twenty unilateral upper limb amputees (age ($M \pm SD$) = 51.40 ± 12.54; 10 amputees with PLP (PLPamp), including four right-arm amputees and amputees without PLP (nonPLPamp), including six right-arm amputees; time since amputation for nonPLPamp and PLPamp ($M \pm SD$) = 23.60 ± 15.49, 18.60 ± 8.55); and 20 matched controls were recruited (see Table 1). Ethical committee approval was received from the Ethical Review Board of the Medical Faculty Mannheim, Heidelberg University, and written informed consent was obtained from all participants. The study protocol adhered to the Declaration of Helsinki.

### *Psychometric assessment of phantom phenomena*

The amputees participated in a psychometric evaluation including a structured interview about the amputation and its consequences [18]. Duration, intensity, and frequency of painful and nonpainful phantom phenomena as well as painful and nonpainful residual limb sensations were assessed by this interview. In addition, the German version of the West Haven–Yale Multidimensional Pain Inventory (MPI; [19,20]) was used in a modified version that separately assessed PLP and residual limb pain [4]. PLP intensity was defined by the MPI phantom pain intensity subscale.

### *Experimental procedure*

We carried out a virtual mirror task in an MR scanner consisting of a virtual reality environment in which participants were seeing a dummy body from a first-person perspective and the scanner bore through MR-compatible goggles (VisuaStimDigital, Resonance Technology, Inc., Northridge, CA, USA). The participants wore a glove on their intact/matched hand capable of tracing movements that were transformed into synchronous movements of both hands of the dummy body in the virtual environment (analogous to the classical mirror box), Figure 1. They were instructed to perform open-close movements with the intact/matched hand at 0.5Hz frequency, paced using auditory tones presented via earphones. They had to focus on the movement of the virtual representation of the moving phantom/mirrored hand [9] and had to try to perceive it as their own hand moving. This set-up created a uniform perception of movement of the phantom hand/intact hand in the amputees and controls and avoided a mixture of imagined and perceived hand movements. To evaluate to what extent the participants felt that the moving phantom/mirrored hand was perceived as their own, we asked them at the end of the MR scan to rate the strength of ownership of the perceived moving hands "How strong was the sensation that you were looking directly at your both hands" (from 0= no sensation to 10=very strong sensation). The protocol comprised alternating 19.8s periods of movement and 19.8s periods of rest, which were repeated six times. The participants were shown if they moved the residual limb along with the intact arm in a training phase and were discouraged from moving the muscles of the residual limb along with the paced open-close hand movements and electromyographic activity of the residual limb was recorded in part of the sample (Brain Products GmbH, Munich, Germany) with a 5000Hz sampling frequency.

--------------------------------------------------------------------------------
----------------INSERT FIGURE 1 AROUND HERE---------
--------------------------------------------------------------------------------

*MRI acquisition and preprocessing*

Echo Planar Images (EPI) were acquired using a Siemens 3 Tesla TRIO scanner (Siemens AG, Erlangen, Germany) in combination with a 12-channel radiofrequency head-coil. In the motor task, 80 volumes were acquired for each participant (2.3x2.3x2.3mm, TR/TE= 3300/45ms, FOV= 220mm, matrix size= 220x220), comprising 40 slices covering the whole brain. A high-resolution 3D

Magnetization Prepared Rapid Gradient image (MPRAGE, 1 mm isotropic voxel, TR/TE=2300/2.98ms) was acquired for anatomical reference.

*Data analysis*

Functional imaging data were analyzed using the FMRIB Software Library (FSL 5.0.9) [21]. For all datasets, motion correction was applied using MCFLIRT [22] and motion-correction parameters were used as nuisance regressors in the design matrix. Spatial smoothing was performed using a 5mm isotropic Gaussian kernel of full-width at half-maximum (FWHM) and high-pass temporal filtering was applied using a Gaussian-weighted least square straight line fitting at 100s cut-off. Registration was performed using a 2-step procedure: EPI images from each scan were first registered to the high resolution T1-weigthed structural image where non-brain structures were removed using Brain Extraction Tool [23]. EPI images were then registered to the standard MNI152 template using 12-parameter affine transformations. The fMRI statistical analysis was carried out using FEAT (fMRI Expert Analysis Tool, Smith 2002). Data from each participant were analyzed separately at a first level of analysis. Trials for the motor task were modeled as a single factor of interest and were convolved with a canonical Gaussian hemodynamic response function, and were entered as a predictor into a general linear model. For both individual and group analyses, areas of significant fMRI responses were determined using clusters identified by a $z> 2.3$ threshold and a Threshold-Free Cluster Enhancement (TFCE)-FWE of $p< 0.05$ [24,25]. Data collected for right-sided (N=10) amputees were mirror-reversed across the mid-sagittal plane before all analyses so that the deafferented hemisphere was consistently aligned. Therefore, in all the following analyses we examined the right hemisphere corresponding to the missing left hand.

We defined a conjunction "phantom/mirror hand ROI" (ROIconj), in accordance to Makin et al. (2013), namely the conjunction of phantom/mirror hand movement-related activation between groups: PLPamp, nonPLPamp and matched controls. For this purpose, a second level analysis was carried out for each group using a mixed-effects analysis as implemented in FLAME (FMRIB's Local

Analysis of Mixed Effects). Then ROIconj was obtained based on the conjunction of the three groups [26] separately within the S1 and M1 using probabilistic maps provided by the Harvard-Oxford structural atlases [27]. Additionally, to relate our findings to Makin et al., (2013) we also examined ROIconj in S1M1, which was defined by the combination of S1 and M1 probabilistic maps.

A second approach consisted in defining an individual "phantom/mirror hand ROI" (ROIind) based on the individual peak voxel statistics, which was used as the center of a spherical ROIs (radius 5 mm). The latter approach was assumed to identify individual and group-related variability in the peak of activation and location of body site representations, which can be expected from previous results on locations of activations in S1 and M1 in amputees with PLP compared to those without PLP and healthy controls (e.g., [4,15]).

Then, we extracted the percent BOLD signal change (%BSC) for each participant and for each ROI (ROIconj, ROIind) in S1, M1 and S1M1 using the FSL Featquery tool [21]. %BSC for each ROI was compared between amputees and controls, and between PLPamp and nonPLPamp using one-way analysis of variance (ANOVA).

We also carried out correlation analyses between PLP intensity and the %BSC extracted from ROIconj and ROIind for S1, M1 and S1M1 in the deafferented side in amputees and the respective side in controls. We also measured Euclidean distances between the ROIind and ROIconj separately for S1, M1, and S1M1 and for each group, and examined the relationship between Euclidean distances and PLP intensity.

For the EMG, we carried out a correlation with task-neural activity during the virtual motor task.

Finally, we compared ownership values of the arm perceived in the mirror between groups (PLPamp, nonPLPamp and controls) using one-way ANOVA.

Threshold for significance was set at $p < .05$. Post-hoc t-test were conducted when necessary, significance threshold was also set at $p < .05$.

---------------------------------------------------------------------------------

----------------INSERT TABLE 1 AROUND HERE---------

---------------------------------------------------------------------------------

**Results**

*Overlap of the task-neural activity between the three groups*

Separate analyses for the PLPamp, nonPLPamp and controls showed neural activity in primary motor and somatosensory cortices, secondary somatosensory cortex, inferior frontal gyrus and lateral occipital cortices in the hemisphere contralateral to amputation (Figure 2A, B, C). Brain areas activated by the virtual mirror task, which overlapped between the three groups, are shown in Figure 2D. There was no significant group difference in task-related activity in the hemisphere ipsilateral to amputation (data not shown).

---------------------------------------------------------------------------------
----------------INSERT FIGURE 2 AROUND HERE---------
---------------------------------------------------------------------------------

Using ROIconj, no significant differences in %BSC were found between amputees (PLPamp and nonPLPamp) and controls using either S1, M1, or S1M1 cortices ($F(1,38) < 3.15$, $p > 0.10$). In addition, no significant differences in %BSC were found between PLPamp and nonPLPamp ($F(1,18) < 0.88$, $p > 0.36$).

Using ROIind, we found that amputees had increased %BSC in S1 ($F(1,38) = 15.2$, $p=0.0004$); M1 ($F(1,38) = 9.98$, $p=0.003$); and S1M1 ($F(1,38) = 13.4$, $p=0.0008$) cortices compared to controls. However, no significant differences in %BSC emerged between between PLPamp and nonPLPamp (($F(1,18) < 0.74$, $p > 0.40$).

There was a complete overlap between ROIconj and ROIind in S1, M1 and S1M1 areas in healthy controls. The location of ROIind for PLPamp and nonPLPamp varied depending on which cortical mask was used (i.e., S1, M1 or S1M1), see Figure 3A, B.

--------------------------------------------------------------------------------
-----------------INSERT FIGURE 3 AROUND HERE---------
--------------------------------------------------------------------------------

*Relationship between neural activity and PLP using ROIconj and ROIind*

Using the entire sample of amputees and ROIconj (as in Makin et al., 2013), no significant relationship was found between PLP intensity and %BSC in S1 ($r=0.13$, $p=0.57$), M1 ($r=0.02$, $p=0.93$) or S1M1 ($r=0.08$, $p=0.71$), see Figure 4A. When we examined the PLPamp group only, we found a significant and linear increase between PLP intensity and %BSC based on ROIconj in S1M1 ($r=0.75$, $p=0.01$) as well as M1 ($r=0.78$, $p=0.0058$), but this did not reach significance for %BSC in S1 ($r=0.54$, $p=0.11$), see Figure 4A.

Using the entire sample of amputees and ROIind, no significant relationship was found between PLP intensity and %BSC in S1 ($r=0.02$, $p=0.93$), M1 ($r=0.10$, $p=0.67$), or S1M1 ($r=0.05$, $p=0.80$). Likewise, for amputees with PLP only, no significant relationship was found between PLP intensity and %BSC in S1 ($r=0.10$, $p=0.93$), M1 ($r=0.10$, $p=0.67$), or S1M1 ($r=0.20$, $p=0.80$), see Fig. 4B. In our sample, there was no significant association between task-neural activity and the EMG signal from the residual limb ($r=-0.01$). Finally, the ownership of the perceived moving hands did not significantly differ between the three groups ($F(1,37)=0.94$, $p=0.40$), (mean± SD= 3.38±0.45 for controls, 3.40±0.64 for nonPLPamp, 4.60±0.64 for PLPamp). Cortical distances between the ROIconj and ROIind for PLPamp and nonPLPamp were not significantly different ($F(1,18)<2.09$, $p>0.17$; see Figure 3 A,B). Data for the controls are not provided since the ROI for the controls completely overlapped with ROIconj.

There was no significant relationship using the whole sample of amputees between PLP intensity and cortical distances between ROIconj and ROind for S1M1 ($r$= .22, $p$ = .36), S1 ($r$= .27, $p$= .27), or M1 ($r$= .32, $p$= .18). Similarly, in the PLPamp subgroup, no significant relationship between PLP intensity and cortical distances between ROIconj and ROind was found ($r \leq .31$, $p \geq .38$).

-----------------------------------------------------------------------------------

----------------INSERT FIGURE 4 AROUND HERE---------

-----------------------------------------------------------------------------------

**Discussion**

We showed that the neural representation of movements of the mirrored hand is different between PLPamp, nonPLPamp and controls. In addition, we found different relationships with PLP intensity, depending on the definition of the regions of interest using either conjunction ROI or individually-defined ROI, using either activation maxima or distance measures and depending on the inclusion of all amputees or only those with phantom pain. Using conjunction ROIs, we did not find a significant association between phantom pain and percent signal change, when we included all amputees. fMRI-related activation during phantom movements only in amputees with PLP showed a positive relationship with PLP intensity using conjunction ROIs for M1 and M1/S1 but not S1 activation. Thus activation in M1 and M1/S1 seems to be related to the intensity rather than the presence of phantom pain per se. Makin et al. (2013) had only two amputees without phantom pain in their sample of 18 amputees, thus the correlation they reported may more reflect an association with the severity of phantom pain rather than the presence of phantom pain per se similar to our results for the ROIconj. When we used individually defined ROIs, we could not find any relationship between phantom pain and percent signal change in S1, M1 or M1/S1, neither when looking at the group of all amputees, nor for the subgroup of the amputees with phantom pain. This is different from Makin et al. (2013) who found that increased neural activity in the S1M1 contralateral to the missing limb was associated with increased PLP. Thus, when the individual differences in the location of the M1/S1 neural activity (ROIind) are taken into account, the relationship between PLP severity and the neural activity in the phantom motor cortex disappeared.

Thus, we suggest that the definition of the ROIs as well as the composition of the amputee sample is important. While both conjoint versus individually-defined ROIs show divergent relationships with PLP intensity, they may also highlight different reorganization processes. ROIconj shows that neural activation in the region formerly representing the intact arm in primary motor and somatosensory cortices is related to PLP intensity. Noteworthy, the conjunction of neural activation between amputees and controls overlapped completely with neural activation in the controls, indicating that the location of ROIconj might be primarily driven by the control group and may encompass only very little of the activation related to phantom movements in the amputees with phantom pain. In fact, one could argue that the activation that goes into the ROIconj analysis, is from those amputees with PLP who are the least reorganized with respect to the location of their maximal activation related to phantom movement.

*Overlap of neural activity between the three groups*

The differences in the location of ROIind compared to ROIconj highlight reorganization processes in the sensory and motor cortices following amputation. In the amputees with PLP, during perception of a moving phantom hand, the location of ROIind shifted depending on which cortical structure was being examined (i.e. S1, M1). In controls, we found a stable cortical activation induced by the virtual mirror task, independent of M1 and S1.

Such neural differences between motor and sensory cortices and between amputees and controls could be related to the different body map representations in primary somatosensory and motor cortices. Indeed, the body maps in M1 are far less fine-grained than the somatotopy found in S1 [28,29]. It has been shown that body maps in M1 reflect movement types rather than individual body sites as in the S1 homunculus [28]. Thus, it is reasonable to assume that a breakdown of lateral inhibition due to deafferentation of a limb affects S1 somatotopy more severely than M1 somatotopy because the deafferentated limb is more broadly and differently represented within M1. In line with this, we found that cortical distances between the peaks of activation between controls and the nonPLPamp were smaller for M1 than for S1, although not significant. The PLPamp, however, seem to show similarly affected somatotopy in both M1 and S1, possibly related to PLP interference of the presence of phantom limb sensations during the virtual motor task [12,30].

*Role of the residual limb during the virtual motor task*

The neural activation we reported in the deafferented hemisphere was not due to associated activity in residual limb muscles since the task we used was implemented to specifically exclude active movements of the residual limb. EMG recordings showed no significant relationship between muscle activity of the residual limb and brain activation in the mirrored movement task. This indicates that – as intended – our task and its induced neural activation were not confounded with muscle activity. Thus, our findings are in line with previous literature using execution of phantom movements [7,11,15,31]. In addition, it enabled us to use a similar task for both amputees and controls, providing therefore homogenous conditions for assessing sensorimotor processes across amputees with and without PLP and controls.

By using a comparable virtual mirror task across amputees and matched controls, we accounted for task-specific alterations but took into account that motor activity related to the amputated limb was absent, which was verified using EMG in a subsample of amputees. In Makin et al. (2013), the amputees were visually instructed to move their phantom, using either motor imagery or executed phantom movements. The latter task is known to actively involve the residual limb, as shown by electromyographic recordings [11,13]. Therefore, the activation measured by Makin et al. (2013) might relate to input from the co-activation of the residual limb representation, which is adjacent to the missing limb representation in S1 and M1 and might have been incorporated in the activation of the ROIconj.

Moreover, we also accounted for subject and group-wise variability in body site representation by systematically using ROIs based on the peak of neural activity and comparing measures derived from motor and somatosensory site cortical representation.

### *Role of execution versus imagery of phantom/mirror hand movements*

The present findings show that although we used a similar task for all amputees, the contribution of M1 and S1 varied between PLPamp, nonPLPamp and controls, supporting differential roles of S1 and M1 for virtual mirror hand movements and in the presence or absence of PLP.

Cortical distances between the peak maxima between the nonPLPamp and the control group were larger in the S1 and smaller in the M1 map. The PLPamp group instead had large cortical distances in both the M1 and S1 maps. Although cortical distances were not significantly related to PLP, the relationship seems to differ depending on whether one examines S1, M1 or S1M1 maps. It would be interesting to carry out similar assessments using a sensory task instead of a motor task and examine how cortical representation might be affected.

In addition, the combination of execution and imagery of phantom motor movements might have resulted in a "blurred" sensorimotor representation of the phantom hand. For instance, motor imagery and motor execution have been shown to rely differentially on primary sensory and motor processing, with the primary sensorimotor cortex being more involved during motor execution and parietal and occipital lobes being more involved during motor imagery [11,32]. Moreover, Raffin et al. showed that phantom movements can be clearly distinguished from imagined phantom movements by demonstrating that amputation leads to a significant slowing down of movement speed as measured behaviorally (mimicking phantom hand movement amplitude and velocity with the intact hand) and psychometrically (questionnaire on mental imagery) [11]. The presence of phantom limb sensations seems also to necessitate more efforts and to lead to longer reaction times when performing a mental rotation task [30]. Further, [11] reported no significant electromyographic activity in residual limb muscles during motor imagery, indicating no incongruent afferent feedback of motor intentions from the residual limb Noteworthy, differences in neural activity during execution and imagery of phantom hand movements could be related to the presence of PLP although this was not investigated by [11]. PLP has been shown to play an important role for the imagery or execution of phantom hand movements. For instance, movements of the elbow and lips showed a shift of neural activity in the former hand representation area, which was related to PLP intensity [15]. Similarly, movements of the lips in PLP showed a shift of the lip representation in the deafferented primary and somatosensory cortices, which was also related to PLP intensity [7]. Using imagery of phantom movements, an increased task-related activity was shown in the hand and in the face cortical area in PLP and not in non-PLP, suggesting co-activation of the mouth and hand representations in primary M1 and S1 in PLP [7]. In addition, using mirrored movements of the intact hand to mimic phantom hand movements, an increased task-related neural activity was reported in M1 and S1 cortices representing the phantom hand in nonPLPamp, but not in PLPamp [8]. Moreover, PLP have been shown to have decreased phantom limb motor control, which was shown by the longer time taken to perform one cycle of phantom movement (stump flexion/extension) compared with non-PLP [11]. Electromyographic activity from the residual limb was also shown to be positively linked to PLP severity [33].

These findings show differences in cortical representation of the phantom hand depending on the task, i.e., imagery or execution of phantom movements, and also between amputees with and without PLP, therefore arguing for not using a common ROI. In addition, all these studies assessed neural activity in both motor and somatosensory cortices, but they did not examine potential differences between the contribution of motor versus sensory processes, which might make sense regarding neural processes involved in phantom movements and /or PLP. Differences in cortical representation might also be influenced whether one examines somatosensory activity related to a sensory stimulation (e.g. tactile) or to motor movement, which follows a different somatotopy.

*Relationship between task-neural activity and PLP intensity*

We reported increased task-related activation in S1 and M1 using ROIind in amputees compared with controls, which was also reported by [7]. In addition, there was no significant difference in task-related activation in S1 or M1 for amputees with and without phantom pain using ROIconj, which is similar to the findings of Makin et al., (2013), when combining task-related activity of all amputees and controls. Current findings also show that although some neural activity remains in the "intact" hand area, the peak of activity in amputees is displaced dorsally and the degree of displacement depends on whether one examines S1 or M1 cortices. The increased task-related activation in amputees is obviously not related to phantom pain since it was present in those with and without pain compared to controls. The higher activation could be related to a high degree of attention to the phantom limb [34] or to the presence of phantom limb sensations, which might lead to more effort related to the task [30]. Only when ROIconj was used, we found a positive association between %BSC and phantom limb pain severity, but not presence for the M1 and M1S1 but not the S1 representation. Specifically, this relationship only emerged within the PLPamp but not the nonPLPamp. Figure 4 shows that there is no mean difference between amputees with and without pain in the ROIconj, but that those with low levels of phantom pain show lower and those with high levels of phantom pain show higher activation in the joint region of interest of all three groups. This increased task-related activation might be indicative of higher excitability related to PLP severity [35] or the persons with more severe pain might have performed the task differently. This relationship with pain severity might also have driven the correlation reported by Makin et al., (2013) who had 16 amputees with and only two without pain. Thus, we suggest that the term "sensorimotor" should be used with caution, especially when related to amputation. We could show that deafferented motor and somatosensory body maps are differently involved in a sensorimotor task. Finally, we focused on alterations in motor and somatosensory pathways following amputation, but we cannot exclude that these alterations might extend to other primary or associative areas such as visual cortex, or temporo-parietal cortex [36].

*Conclusion*

We found that motor and sensory cortical areas show different activation patterns for amputees and healthy controls dependent on the definition of the region of interest. When individual ROIs were used, the amputees showed higher activation in M1 and M1S1 than when joint ROIs were used. M1 and S1M1 activation showed a relationship with PLP severity only in the PLPamp group and only when joint ROIs were used. However, the origin of this relationship is not clear.

Longitudinal studies are needed to examine how functional reorganization differs between S1 and M1 in terms of extent, peak activation and speed, and its relationship with PLP. A better understanding of the role of M1 or S1 in PLP could help to optimize the definition of neural sites to be targeted using techniques such as transcranial magnetic stimulation or neurofeedback.

**Acknowledgments**

This study was supported by a grant from the Deutsche Forschungsgemeinschaft (SFB1158/B07) to H.F. and J.A. and a European Research Council Advanced Grant (No. 230249) to H.F. We would like to thank Astrid Wolf for help with recruitment.

**Competing Interests statement**

This manuscript reflects only the author's views and the funding agencies are not liable for any use that may be made of the information contained therein. The authors declare no conflict of interest.

**Author Contributions statement**

*Study design:* Andoh J, Milde C, Diers M, Bekrater-Bodmann R, Trojan J, Fuchs X, Flor H

*Data analysis:* Andoh J and Milde C

*Data interpretation:* Andoh J, Milde C, Diers M, Bekrater-Bodmann R, Trojan J, Fuchs X, Becker S, Desch S, Meijer O, Flor H

*Manuscript writing:* Andoh J, Milde C, Flor H

*Manuscript revision:* Andoh J, Milde C, Diers M, Bekrater-Bodmann R, Trojan J, Fuchs X, Becker S, Desch S, Meijer O, Flor H